\documentclass{emulateapj}
\usepackage{amsmath}

\def\stacksymbols #1#2#3#4{\def\theguybelow{#2}
        \def\verticalposition{\lower#3pt}
        \def\spacingwithinsymbol{\baselineskip0pt\lineskip#4pt}
        \mathrel{\mathpalette\intermediary#1}}
\def\intermediary #1#2{\verticalposition\vbox{\spacingwithinsymbol
        \everycr={}\tabskip0pt
        \halign{$\mathsurround0pt#1\hfil##\hfil$\crcr#2\crcr
                \theguybelow\crcr}}}

\shorttitle{COSMIC RAY-DOMINATED AGN JETS}
\shortauthors{GUO \& MATHEWS}

\begin{document}
\bibliographystyle{apj} 

\title {Cosmic Ray-Dominated AGN Jets and the Formation of X-ray Cavities in Galaxy Clusters}

\author{Fulai Guo\altaffilmark{1} and William G. Mathews \altaffilmark{1}}

\altaffiltext{1}{UCO/Lick Observatory, Department of Astronomy and Astrophysics, University of California, Santa Cruz, CA 95064, USA; fulai@ucolick.org}

\begin{abstract}
It is widely accepted that feedback from active galactic nuclei (AGN) plays a key role in the evolution of gas in groups and clusters of galaxies. Unequivocal evidence comes from quasi-spherical X-ray cavities observed near cluster centers having sizes ranging from a few to tens of kpc, some containing radio emission. Cavities apparently evolve from the interaction of AGN jets with the intracluster medium (ICM).  However, in numerical simulations it has been difficult to create such fat cavities from narrow jets. Ultra-hot thermal jets dominated by kinetic energy typically penetrate deep into the ICM, forming radially elongated cavities at large radii unlike those observed. Here, we study very light jets dominated energetically by relativistic cosmic rays (CRs) with axisymmetric hydrodynamic simulations, investigating the jet evolution both when they are active and when they are later turned off. We find that, when the thermal gas density in a CR-dominated jet is sufficiently low, the jet has a correspondingly low inertia, and thus decelerates quickly in the ICM. Furthermore, CR pressure causes the jet to expand laterally, encounter and displace more decelerating ICM gas, naturally producing fat cavities near cluster centers similar to those observed. Our calculations of cavity formation imply that AGN jets responsible for creating fat X-ray cavities (radio bubbles) are very light, and dominated by CRs. This scenario is consistent with radio observations of Fanaroff-Riley I jets that appear to decelerate rapidly, produce strong synchrotron emission and expand typically at distances of a few kpc from the central AGN.

\end{abstract}


\section{Introduction}
\label{section:intro}

Feedback from active galactic nuclei (AGNs) is increasingly recognized to play an important role in the evolution of elliptical galaxies, groups and clusters of galaxies. It is considered to be a prime candidate for solving the ``overcooling" problem associated with galaxies at the bright end of the galaxy luminosity function \citep{croton06,teyssier10}, i.e., cosmological simulations without AGN feedback predict massive galaxies that are too bright and too blue when compared to massive galaxies in the nearby universe (\citealt{kravtsov05}; \citealt{borgani09}). Massive galaxies sit at the centers of large dark matter haloes, and the ``overcooling"problem is thus related to the so-called ``cooling flow" problem in galaxy clusters, groups, and giant elliptical galaxies. High-resolution X-ray observations from the {\it Chandra} and {\it XMM-Newton} Telescopes have shown that, though the gas in these systems emits copiously in X-rays, the amount of gas cooling to very low temperatures is much less than the amount of hot gas with a short cooling time (see \citealt{peterson06} for a recent review). This suggests that some heating mechanism is at work to counterbalance radiative losses. It is now widely agreed that AGN feedback plays a key role in heating the hot gas in these systems (see \citealt{mcnamara07} for a recent review).

The strongest observational evidence for AGN feedback is in galaxy clusters where high-resolution X-ray observations have detected numerous X-ray-deficient cavities, some of which are associated with radio jets and spatially coincident with radio bubbles/lobes \citep{mcnamara07}. Most cavities have spatial sizes of a few to a few tens kpc and are located near cluster centers with cluster-centric distances of a few to a few tens kpc \citep{mcnamara07, diehl08}. These cavities are roughly spherical or elongated in the tangential direction (i.e., perpendicular to the radial jet axis), and are likely rising buoyantly in the intracluster medium (ICM). Observations indicate that roughly $70$ - $75\%$ of X-ray bright cool-core clusters harbor detectable cavities \citep{dunn05}, and the associated AGN energies are sufficient or nearly sufficient to stop the gas from cooling in most clusters containing detectable cavities \citep{rafferty06}. This suggests that AGN feedback indeed plays a significant role in the thermodynamical evolution of galaxy clusters \citep{guo08b}. X-ray cavities have also been detected in elliptical galaxies \citep{allen06} and in galaxy groups \citep{dong10}, but it is usually more difficult to detect cavities in smaller halos with fainter X-ray emission.

Combined radio and X-ray observations of X-ray cavities (e.g., \citealt{dunn05}; \citealt{mcnamara09}) suggest that these cavities are inflated by bipolar jets emanating from AGNs in cluster-central galaxies. It is natural to study AGN feedback by investigating the evolution of AGN jets in these systems. In recent years, numerical simulations of jets in galaxy clusters have been conducted by many authors \citep[e.g.,][]{omma04,zanni05,vernaleo06,brueggen07,oneill10,morsony10}, who mainly focus on the impact of AGN outbursts on the thermal and dynamical evolution of the ICM. In contrast, there are relatively few gas-dynamical studies of the physics of AGN outbursts itself or the content (composition) of AGN jets that directly affect the impact of AGN feedback on the ICM. In most of these calculations, it is  assumed that AGN jets contain ultra-hot thermal gas with jet-to-ICM density contrasts $\eta=\rho_j/\rho \approx 0.01$. The energy fluxes of these thermal jets are commonly dominated by the kinetic energy. Simulations indicate that these jets usually penetrate deep into the ICM, forming radially-elongated cavities at large radii (e.g., \citealt{reynolds02}, \citealt{omma04}, \citealt{zanni05}, \citealt{vernaleo06},\citealt{oneill10}), but never forming the ``fatter" quasi-spherical X-ray cavities observed near cluster centers. This `cavity formation' problem suggests that the adopted model for either AGN jets or the ICM has at least one inaccurate key ingredient, which may lead to quantitively or even qualitatively incorrect results.

The cavity formation problem has been previously noticed by \citet{sternberg07}, who propose wide, conical jets with typical half-opening angles $\alpha \gtrsim 50^{\circ}$ to form fat cavities in galaxy clusters. Rapidly precessing jets  with large precessing angles can also produce quasi-spherical fat cavities (\citealt{sternberg08a}; but slowly precessing jets may produce multiple pairs of cavities, \citealt{falceta10}). Alternatively, jet simulations by \citet{brueggen07} suggest that narrow jets could result in fat cavities if significant random motions are present in the ICM. Fat cavities are seen in simulations of \citet{morsony10}, which include both jet precession and ICM random motions. However, it is unclear if any of the above three conditions is common in extragalactic AGN jets responsible for forming fat cavities. Consider for example the Virgo cluster, the closest galaxy cluster which exhibits detailed interactions of AGN feedback with the ICM (e.g., \citealt{million10}). The $90$ cm radio image of M87 by \citet{owen00} shows two regular, nearly spherical radio lobes, unlikely to have been strongly shaped by random gas motions which tend to form irregular bubbles (see Figure 1 in \citealt{brueggen07}). Multi-wavelength observations of the famous M87 jet clearly show that it is highly collimated even on parsec scales (\citealt{junor99}; \citealt{kovalev07}). Narrow radio jets have also been observed in numerous other systems, indicating that the jet opening angle is generally very small. If the eastern ear-shaped radio lobe is formed by the jet traced by the eastern radio arm (see Figure 2 in \citealt{owen00}), the well-collimated eastern arm may suggest that the jet does not have a large precession angle as required by \citet{sternberg08a} to create fat bubbles. Observations of M87 thus motivate us to explore other missing ingredients of AGN feedback, which may be responsible for forming fat cavities.

There is indeed one important ingredient of AGN feedback missing in previous simulations -- the relativistic cosmic ray (CR) component of AGN jets. CRs provide pressure with negligible inertia, which can puff up jets to form distended cavities and radio lobes near cluster centers. The same effect can also be achieved by thermal jets with very low density contrasts and high internal pressures, but thermal particles in these jets are actually relativistically hot (see discussions in Section~\ref{section:why}). Furthermore, numerous detections of radio jets and lobes associated with X-ray cavities point toward the existence of a significant CR electron population. Deep radio observations of Fanaroff-Riley type I (FR I) radio jets, which are likely to be responsible for creating X-ray cavities, indicate that these jets decelerate rapidly and produce strong synchrotron emission (flaring) at typical distances of a few kpc from central nuclei (e.g., \citealt{laing06}; \citealt{laing08}). For the first time, in this paper we perform numerical simulations of CR-dominated jets and follow the dynamical evolution of these jets in a typical galaxy cluster. These jets contain a dominant CR component and a very small thermal component. They are very light and yet have a large energy flux and high internal pressure. Due to the reduced jet inertia and momentum with enhanced lateral expansion driven by CR pressure, these jets naturally produce fat cavities near cluster centers, resembling observed cavities very well. 

Radio jets have been previously proposed to explain extended (particularly FR II) extragalactic radio sources \citep{longair73,scheuer74,blandford74}. The propagation of purely relativistic radio jets has been studied with conservation-law calculations \citep{bicknell94,laing02}, but numerical simulations usually focus on jets composed of ideal thermal gas (some including magnetic fields or relativistic bulk motions, e.g., \citealt{perucho07}, \citealt{oneill10}). Unlike \citet{scheck02} and \citet{perucho07}, who adopt an equation of state for a thermalized relativistic ideal gas, our hydrodynamical calculations follow an additional relativistic CR component that can have a power-law energy distribution as expected from shock acceleration.  We find that fat X-ray cavities observed in galaxy groups and clusters are most naturally formed from jets in which the CR energy density dominates and which are very light, i.e. containing non-relativistic thermal gas with densities that are $\sim10^{-4}$ times the central cluster gas density.  Such light jets have also been considered for example by \citet{krause03} and \citet{gaibler09} in studies of purely thermal FR II jets and cocoons. For the first time, we numerically investigate the evolution of two-fluid jets composed of both thermal gas and CRs.

The rest of the paper is organized as follows. In Section~\ref{section2}, we describe the physics of galaxy clusters with CRs, and numerical setup. We present our results, in particular, how CR-dominated jets form fat cavities, in Section~\ref{section:results}, and summarize our main conclusions with implications in Section~\ref{section:conclusion}.
 
   \begin{figure*}
\plottwo{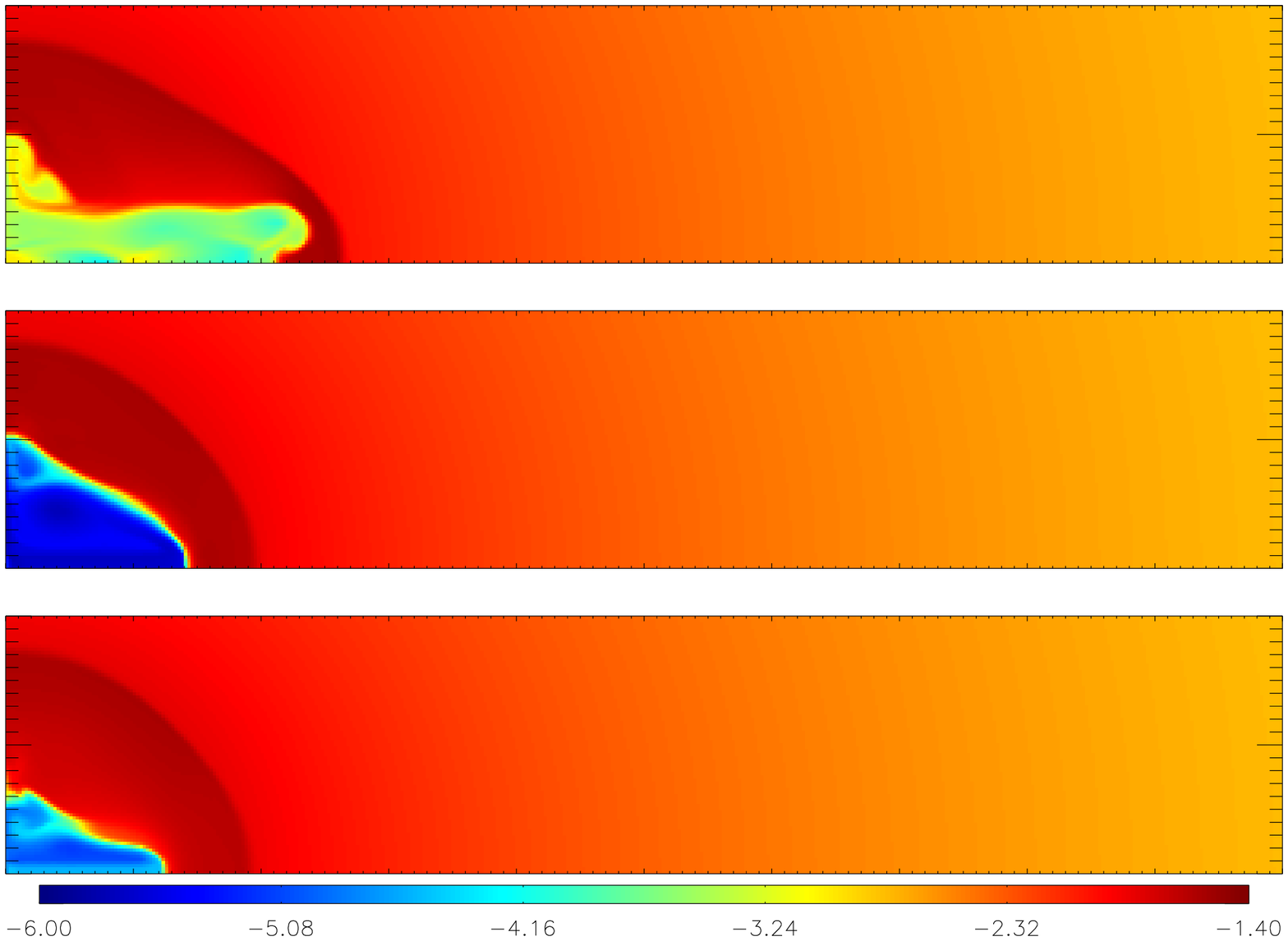}{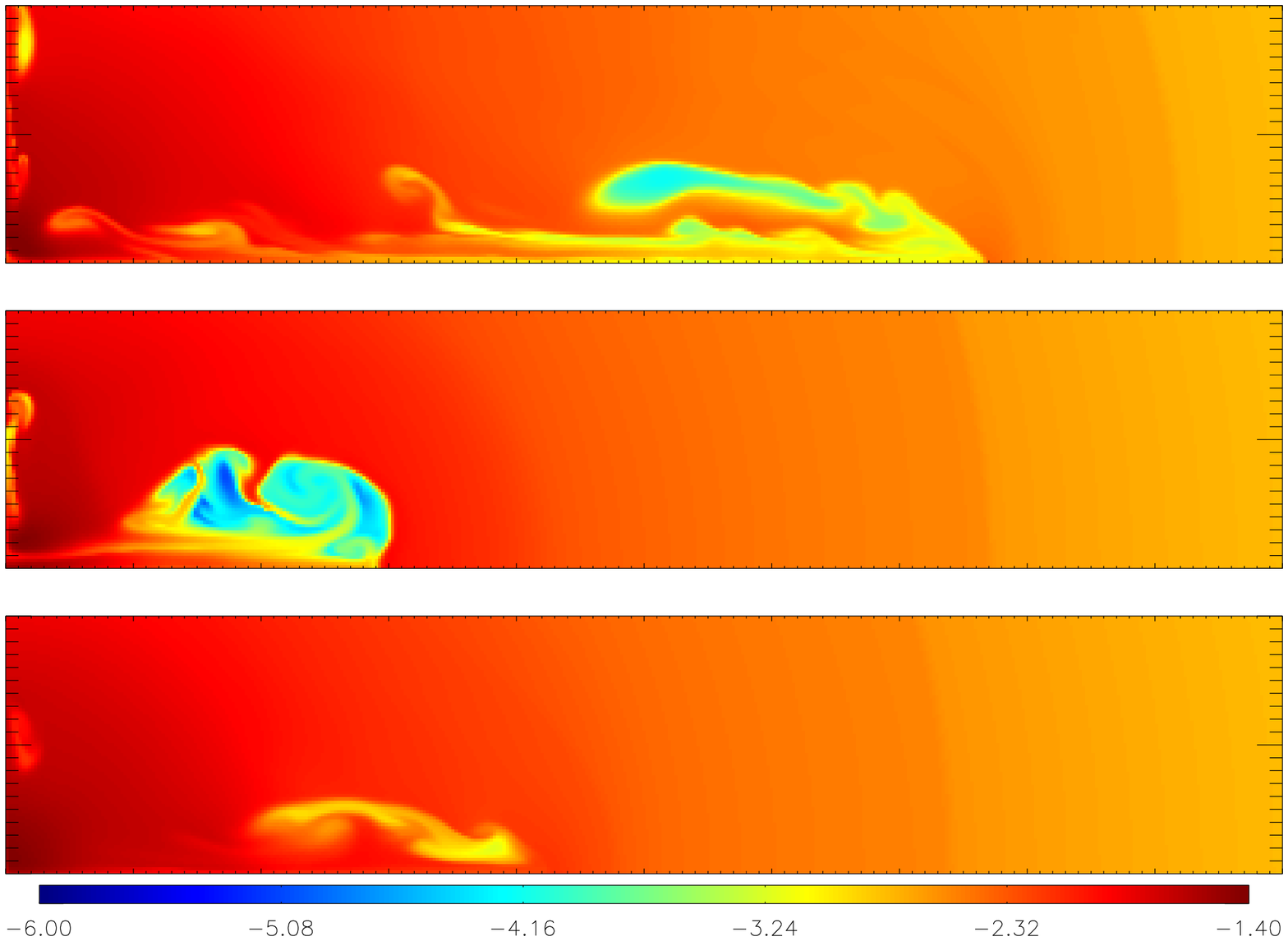}
\caption{Central slices ($100\times20$ kpc) of  log $(n_{\rm e}/{\rm cm}^{-3})$ in runs J0 (top panels), J1 (middle panels) and J1-A (bottom panels) at time $t=10$ Myr (left panels) and $t=70$ Myr (right panels). The horizontal and vertical axes represent $z$ and $r$ respectively.   }
 \label{plot1}
 \end{figure*} 
 
\section{Equations and Numerical Methods}
\label{section2}

\subsection{Cluster Physics with Cosmic Rays}
\label{section:equation}

X-ray cavities are usually thought to be inflated by bipolar jets emanating from AGNs located at cluster centers. 
Radio synchrotron emission has been detected from many jets and X-ray cavities, confirming the existence of a significant CR component, which may be transported along the jets and/or created in strong shocks as the jets encounter the ICM. The combined hydrodynamic evolution of thermal gas and CRs can be described by the following four equations:

\begin{eqnarray}
\frac{d \rho}{d t} + \rho \nabla \cdot {\bf v} = 0,\label{hydro1}
\end{eqnarray}
\begin{eqnarray}
\rho \frac{d {\bf v}}{d t} = -\nabla (P+P_{\rm c})-\rho \nabla \Phi ,\label{hydro2}
\end{eqnarray}
\begin{eqnarray}
\frac{\partial e}{\partial t} +\nabla \cdot(e{\bf v})=-P\nabla \cdot {\bf v}
   \rm{ ,}\label{hydro3}
   \end{eqnarray}
\begin{eqnarray}
\frac{\partial e_{\rm c}}{\partial t} +\nabla \cdot(e_{\rm c}{\bf v})=-P_{\rm c}\nabla \cdot {\bf v}+\nabla \cdot(\kappa\nabla e_{\rm c})+\dot{S_{\rm c}}
   \rm{ ,}\label{hydro4}   \end{eqnarray}
  \\ \nonumber
\noindent
where $d/dt \equiv \partial/\partial t+{\bf v} \cdot \nabla $ is the
Lagrangian time derivative, $P_{\rm c}$ is the CR pressure, $e_{\rm
  c}$ is the CR energy density, $\kappa$ is the CR diffusion
coefficient, $\dot{S_{\rm c}}$ is the CR source term, and all other variables have their usual
meanings. Pressures and energy densities are related via
$P=(\gamma-1)e$ and $P_{\rm c}=(\gamma_{\rm c}-1)e_{\rm c}$, where we
assume $\gamma=5/3$ and $\gamma_{\rm c}=4/3$. We ignore radiative cooling, which is unimportant during our short-duration ($\lesssim 100$ Myr) simulations of cavity formation. The gas temperature is related to the gas pressure and density via the ideal gas law:
\begin{eqnarray}
T=\frac{\mu m_{\mu}P}{k_{\rm B} \rho} {\rm ,}
   \end{eqnarray}
where $k_{\rm B}$ is Boltzmann's constant, $m_{\mu}$ is the atomic mass unit, and $\mu=0.61$ is the molecular weight. To avoid confusion we denote the gas pressure $P$ as $P_{\rm g}$ in the rest of the paper.
  
Our model and methods are generally applicable to all reasonably relaxed clusters,
but for concreteness, we adopt simulation parameters appropriate for the well-observed nearby 
Virgo cluster. The initial radial profiles of gas density and temperature are taken from the analytic fits to the observations provided by \citet{ghizzardi04}. We choose a gravitational field that establishes exact hydrostatic equilibrium for the initial gas pressure profile. At time $t=0$ we assume that the CR energy density is zero in the cluster gas, $e_{\rm cr}=0$, but at later times CRs enter the cluster in jets. Equation \ref{hydro4} describes the CR advection and diffusion in the ICM. The CR diffusion coefficient
$\kappa$ is poorly known but may vary inversely with the gas
density since the magnetic field is probably larger in denser gas
\citep{dolag01}. We adopt the following functional dependence of the diffusion
coefficient on the electron number density $n_{\rm e}$:
\begin{eqnarray}
\kappa=
\begin{cases}
10^{29}(n_{{\rm e}0}/n_{\rm e}) \text{~cm}^{2} \text{~s}^{-1}    & \quad \text{when } n_{\rm e}> n_{{\rm e}0} \text{,}\\
10^{29} \text{~cm}^{2} \text{~s}^{-1}    & \quad \text{when } n_{\rm e} \leq n_{{\rm e}0} \text{,}
\end{cases}
\end{eqnarray}
\noindent
where $n_{{\rm e}0}=10^{-5}$ cm$^{-3}$. This level of diffusion does not strongly affect the early evolution of cavities, and our results are fairly insensitive to it. During their diffusion, CRs interact with magnetic irregularities and Alfv\'{e}n waves, effectively exerting CR pressure gradients on the thermal gas (equation \ref{hydro2}). Pressure gradients in the CR component act directly on the gas by means of magnetic fields frozen into the gas but the magnetic energy densities are too small to contribute to the gas dynamics and do not appear in the equations above. Recent observations also suggest that the magnetic field is likely to be dynamically subdominant in X-ray cavities, where the energy density is mainly contributed by a particle population that is not responsible for the observed synchrotron emission \citep{croston08,hardcastle10,dunn10}. We neglect other more complicated (probably secondary) interactions of CRs with thermal gas, e.g., Coulomb interactions, hadronic collisions, and hydromagnetic-wave-mediated CR heating, that depend on the CR energy spectrum and may provide additional heating effects for the ICM (e.g., \citealt{guo08a}). We also neglect CR energy losses from synchrotron emission; although synchrotron losses may be important for high-energy CR electrons, most of the contribution to the integrated CR energy density $e_{\rm c}$ comes from low-energy CR electrons and possibly CR protons.

Our method of CR modeling has been successfully used in a series of papers, e.g., \citet{mathews08a}, \citet{mathews08}, \citet{mathews09}, \citet{guo10a}, and \citet{guo10b}, to which the reader is referred to for further details. When X-ray cavities were formed in these previous studies, we ignored the complex jet physics and assumed that CRs were injected {\it in situ} into a small region slightly offset from the cluster center. Once formed, X-ray cavities rise buoyantly in the ICM. In this paper, we take a step further and study in detail how CR-dominated jets form cavities in the ICM. To this end, we assume that the CR source term in Equation \ref{hydro4} is $\dot{S_{\rm c}}=0$, and all the CRs are injected into the ICM by the jet at the inner boundary of our simulations, as explained in the following subsection.
 
\subsection{Numerical Setup and Jet Injection}
\label{section:jetinjection}

Equations (\ref{hydro1}) $-$ (\ref{hydro4}) are solved in $(r, z)$
cylindrical coordinates using a two-dimensional Eulerian code similar
to ZEUS 2D \citep{stone92}. Our code contains a background gravitational potential, CR diffusion, and an 
energy equation for CRs. The computational grid consists of $400$ equally
spaced zones in both coordinates out to $100$ kpc plus additional
$100$ logarithmically-spaced zones out to $1$ Mpc. For both thermal and CR fluids, we adopt outflow boundary conditions at the outer boundary and reflective boundary conditions at the inner boundary except for those inner boundaries representing the jet base when the jet is active.
The jet inflow is introduced along the $z$-axis through a nozzle placed at the cluster center.
At its base, the jet has a radius of $r_{\rm jet}=1$ kpc ($4$ zones along the $r$ axis) and is initialized by inflow 
boundary conditions similar to those in \citet{brueggen07}. By preparing the jet properties before it enters the computational grid, we avoid an initial entrainment of cluster gas into the jet as usually occurs if the jet is initialized within active zones. 

The jet contains both thermal gas and CRs and is injected with cylindrical geometry having an initial opening angle of $0$ degrees. At their base in central ghost zones jet inflows are described by six parameters: the density ratio $\eta=\rho_{\rm j}/\rho_{0}$ of thermal gas within the jet to the ambient gas at the cluster center, the energy density of thermal gas within the jet $e_{j}$, the CR energy density within the jet $e_{ \rm jcr}$, the jet velocity $v_{\rm jet}$, the jet radius $r_{\rm jet}$, and the jet duration $t_{\rm jet}$. The jet power can then be written as
\begin{eqnarray}
P_{\rm jet}=P_{\rm ke}+P_{\rm cr} +P_{\rm th}{\rm ,}
   \end{eqnarray}
where $P_{\rm ke}=\rho_{j} v_{\rm jet}^{3}\pi r_{\rm jet}^{2}/2$ is the jet kinetic power, $P_{\rm cr}=e_{ \rm jcr} v_{\rm jet}\pi r_{\rm jet}^{2}$ is the jet CR power, and $P_{\rm th}=e_{ \rm j} v_{\rm jet}\pi r_{\rm jet}^{2}$ is the jet thermal power. The total energy injected by the jet can be written as $E_{\rm jet}=P_{\rm jet}t_{\rm jet}$. In the computations presented in this paper, we take $v_{\rm jet}=0.05c$, where $c$ is the speed of light, $r_{\rm jet}=1$ kpc, and $t_{\rm jet}=10$ Myr (except that in run J3, $t_{\rm jet}=2.72$ Myr to keep $E_{\rm jet}$ the same as in our main run J1). The jet has a Mach number of $\sim 23$ with respect to the initial cluster gas at the origin. AGN jets on parsec scales likely have relativistic velocities, but they may undergo rapid deceleration on kpc scales \citep{laing06}. Our initial jet speed $v_{\rm jet}=0.05c$ represents jets that have already gone through the initial deceleration stage. We have also run simulations with $v_{\rm jet}=0.1c$ or $0.2c$, and found qualitatively similar results. The jet internal thermal energy density $e_{j}$ is determined by $\eta$ and the jet thermal temperature $T_{\rm j}$, which is chosen to be $100$ times the initial ambient gas temperature at the origin. Since the jet power in our simulations is dominated by either the kinetic power or the CR power, our results are not sensitive to the value of $T_{\rm j}$ unless it is so high that $e_{\rm j} \gtrsim e_{\rm jcr}$. The jet power in our main run J1 is $P_{\rm jet}\sim 1.3\times 10^{44}$ erg s$^{-1}$, which corresponds to a powerful FR I radio source and is chosen to roughly produce X-ray cavities with radii $\sim 10$ kpc near the center of our fiducial cluster. The jet energy and power are consistent with typical values estimated from cavity energetics, which span a large range \citep{birzan04}. In observation, the AGN mechanical luminosity is usually averaged within the cavity rising time (the buoyancy timescale), which is likely much longer than the jet duration. See Table \ref{table1} for the jet parameters and energetics in all runs presented in this paper. 

\begin{table}
 \centering
 \begin{minipage}{80mm}
  \renewcommand{\thefootnote}{\thempfootnote} 
  \caption{List of Simulations}
    \vspace{0.1in}
  \begin{tabular}{@{}lccccccc}
  \hline Run& $\eta$& {$e_{\rm jcr}$\footnote{The initial CR energy density at the jet base (in units of $10^{-9}$ erg cm$^{-3}$).}} &
         {$E_{\rm cr}$\footnote{$E_{\rm cr}$, $E_{\rm ke}$, $E_{\rm th}$, and $E_{\rm jet}$ are, respectively, the injected CR, kinetic, thermal, and total energy (in units of $10^{58}$ erg) by the jet during its active phase $t\leq t_{\rm jet}$ (e.g., $E_{\rm cr}=P_{\rm cr}t_{\rm jet}$).}} & {$E_{\rm ke}$\footnotemark}&{$E_{\rm th}$\footnotemark}&{$E_{\rm jet}$\footnotemark}&{$P_{\rm jet}$\footnote{The jet power in units of $10^{44}$ erg s$^{-1}$.}}\\  \hline J0 &0.01&0&0& 3.13 &1.07 &4.21&1.3\\ J1 & 0.0001&$2.96$  & 4.17&0.03&0.01&4.21&1.3\\ J1-A & 0.0001&$0$&
       0&  0.03 & 0.01 &0.04 &0.01\\ J2 & 0.0001&$1.52$&
        2.14&0.03 & 0.01 &2.18 &0.69\\  {J3\footnote{In run J3, the jet is active for a duration of $t_{\rm jet}=2.72$ Myr (to keep $E_{\rm jet}$ the same as in run J1), while in all other runs, $t_{\rm jet}=10$ Myr.}} & 0.01&$8.0$&
        3.06&0.85 & 0.29 &4.21 &4.9\\ 
         
          \hline
\label{table1}
\end{tabular}
\end{minipage}
\end{table}

\section{Results}
\label{section:results}

\subsection{Kinetic Energy-dominated Jets}

Numerical simulations suggest that jets dominated by kinetic energy usually penetrate easily into the ICM, forming low-density cavities at cluster centric distances much larger than those in real clusters observed at similar ages (e.g., \citealt{reynolds02}, \citealt{omma04}, \citealt{zanni05}, \citealt{vernaleo06}, \citealt{oneill10}). Furthermore, these jet-inflated cavities are strongly elongated in the (radial) jet direction, while most observed cavities are quite spherical, or elongated in the tangential direction (e.g., the cavities in the Perseus cluster, \citealt{fabian06}). The goal of this paper is to compute the dynamical evolution of CR-dominated jets and investigate if these physically motivated jets can form fat X-ray cavities near cluster centers.

We first present our control run J0, which follows the evolution of a typical kinetic energy-dominated jet. As listed in Table \ref{table1}, the parameters of run J0 are $\eta=0.01$, and $e_{\rm jcr}=0$, similar to those often adopted by previous authors (e.g., \citealt{reynolds02}, \citealt{vernaleo06}, \citealt{oneill10}). As usual, we assume that the jet thermal temperature at the jet base is $100$ times the initial cluster gas temperature at the origin. Thus the jet is initially in pressure equilibrium with the ambient gas. This assumption is the most used in jet simulations, but may not hold for supersonic jets moving down the cluster pressure gradient. Furthermore, the initial jet pressure is determined by physical processes directly associated with the central supermassive black hole, and is unlikely to be equal to the pressure of central hot ICM. The shock-heating of the jet material and the production of CRs within the jet may make the jet over-pressured. We will explicitly explore the effect of jet pressure on the formation of fat cavities in Subsection \ref{section:why}.

\begin{figure}
\plotone{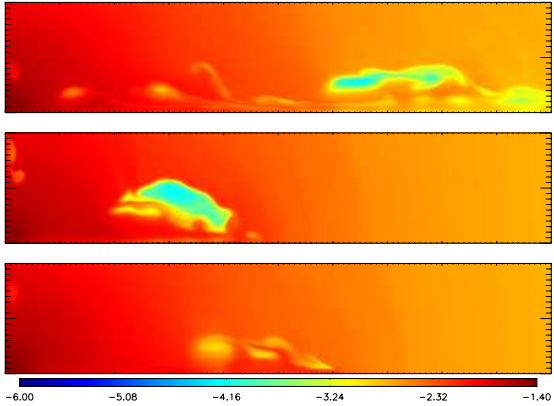}
\caption{Same as Figure \ref{plot1}, but at $t=100$ Myr. }
 \label{plot2}
 \end{figure}

Figure \ref{plot1} shows central 2D slices ($100$ $\times$ $20$ kpc) of electron number density in logarithmic scale in runs J0 (top panels), J1 (middle panels) and J1-A (bottom panels) at time $t=10$ Myr (left panels) and $t=70$ Myr (right panels). The same images at $t=100$ Myr are shown in Figure \ref{plot2}. As clearly seen in these figures, the kinetic-energy-dominated jet in run J0 penetrates easily into the ICM, and reaches $\sim 100$ kpc at $t=100$ Myr. The jet creates a low-density cavity, which is strongly elongated in the jet direction. These features have been previously seen in many simulations of kinetic-energy-dominated jets, and are clearly inconsistent with observations of most X-ray cavities, which are often approximately spherical and appear to rise buoyantly at cluster centric radii $\lesssim 50$ kpc in times $t \lesssim 100$ Myr \citep{mcnamara07}.

  \begin{figure*}
\plotone{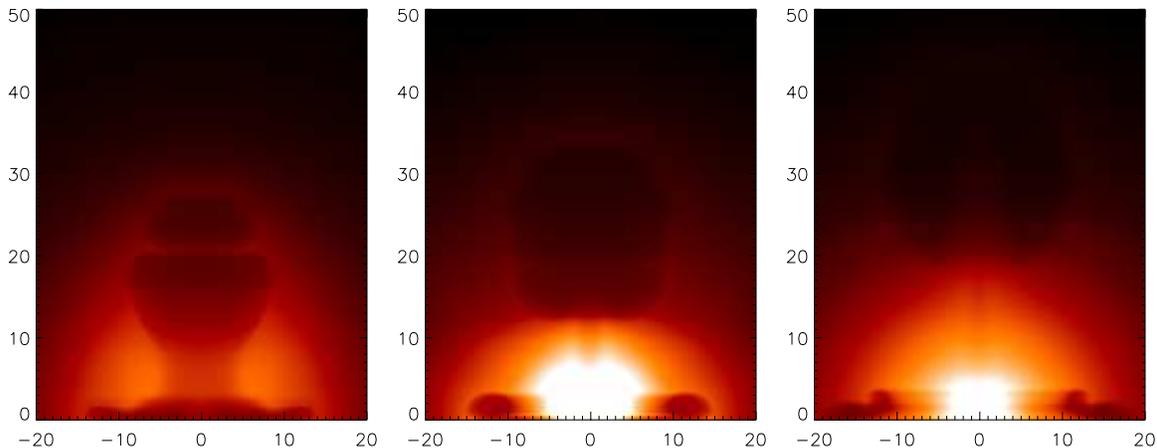} 
\caption{Central regions ($50\times40$ kpc) of  synthetic X-ray surface brightness maps (line of sight integrated projections of the cooling rate $n_{\rm e}n_{\rm i}\Lambda(T,Z)$ perpendicular to the jet axis) for run J1 at $t=60$ (left panel), 80 (middle panel) and 100 Myr (right panel). The cavity is clearly seen as it rises in the ICM.}
 \label{plot3}
 \end{figure*} 
 
 \subsection{Cosmic Ray-dominated Jets}
 
In run J0, the jet injects into the ICM a total energy of $E_{\rm jet}=4.21\times10^{58}$ erg, which is mainly kinetic. Our preferred jet J1 has the same amount of total jet energy, but it is dominated by CRs. We assume that the jet in run J1 has the same initial temperature $T_{\rm j}$ and velocity $v_{\rm jet}$ as in run J0, but has a much smaller gas density ($\eta=10^{-4}$) and consequently a very small kinetic energy. We choose the initial CR energy density in the jet to be $e_{\rm jcr}=2.96\times 10^{-9}$ erg cm$^{-3}$ so that  $E_{\rm jet}$ in run J1 is the same as that in run J0 (see Table \ref{table1} for the jet energetics). To emphasize the role of CRs in the jet evolution, we performed another run J1-A, which is exactly the same as run J1 except that the jet in this run contains no CRs ($e_{\rm jcr}=0$). For reference, the central electron number density and temperature in the Virgo cluster are $n_0 \sim 0.1$ cm$^{-3}$ and $T_{0} \sim 1.6$ keV respectively \citep{ghizzardi04}.
 
In Figures \ref{plot1} and \ref{plot2}, we see a striking difference on the evolution of jets J0 and J1: Unlike the jet in run J0, which forms a radially-elongated cavity at large radii, the CR-dominated jet in run J1 forms a `fat' low-density cavity near the cluster center. Due to its very low inertia and momentum, the light CR-dominated jet in run J1 decelerates much faster, and reaches a smaller distance than the thermal jet in run J0 at a same time, as seen in the left panel of Figure \ref{plot1}, which also shows that the CR-dominated jet inflates a wider cavity which is expanded laterally due to the CR pressure accumulated within the shorter jet and cavity. As the jet turns off at $t_{\rm jet}=10$ Myr, the expanded jet head decelerates quickly in the ICM due to its low inertia, and stops moving outward at $t\sim 20$ Myr. As the jet nearly transfers all its initial momentum to the surrounding gas, it becomes indistinguishable from the cavity, which rises slowly outward due to buoyancy. In contrast, the thermal jet in run J0 has a much higher inertia and continues to move outward due to its initial momentum even at the end of the simulation ($t=100$ Myr). 

At $t=70$ Myr, the jet-inflated cavity in run J1 is located at a distance of $\sim 20$ kpc from the cluster center, having a radius of $\sim 10$ kpc. As the cavity rises in the radial direction, relatively cooler thermal gas converges beneath the cavity toward the jet axis, forming a radial filament or `thermal jet' which penetrates outward through the cavity. This filament is visible at $t=70$ Myr (the right-middle panel of Figure \ref{plot1}) and $100$ Myr (the middle panel of Figure \ref{plot2}). This thermal jet has been seen in many previous simulations of X-ray cavities (e.g., \citealt{churazov01}; \citealt{reynolds05};  \citealt{sternberg08}), and is denoted as `cavity jet' by \citet{mathews08a} or `drift' by \citet{pope10}. Outward moving radial streams of cool, low-entropy gas are a natural post-cavity gas-dynamical development that probably explains cold, nearly radial filaments commonly seen in cool core clusters. 

Figure \ref{plot3} shows central regions ($50\times40$ kpc) of synthetic X-ray surface brightness maps for run J1 at three different times $t=60$ (left panel), $80$ (middle panel), and $100$ Myr (right panels). These X-ray maps are produced by integrating the radiative cooling rate $n_{\rm e}n_{\rm i}\Lambda(T,Z)$ \citep{sd93} along a direction perpendicular to the jet axis (plane of the sky). Here $n_{\rm i}$ is the total number density of ions and we take an average metallicity of $Z=0.4$. In Figure \ref{plot3}, X-ray cavities with radius $\sim 10$ kpc are clearly seen. Interestingly, in the left and middle panels, the cavity is enclosed by a rim or shell significantly brighter than the ambient gas, a feature often seen in real X-ray cavities (e.g., \citealt{fabian2000}; \citealt{nulsen02}). Dense thermal shells form naturally as the cavity is created, expands in the ICM and may become colder than the ambient gas, as suggested by observations.

In Figures \ref{plot1} - \ref{plot3}, we notice that a small fraction of CR-filled low density gas rises up in directions perpendicular to the jet axis. As discussed in the next subsection, these small cavity-like features are produced by backflowing cocoon gas which reaches the $z=0$ plane and subsequently rises in the $r$ directions, forming small (toroidal) cavities perpendicular to the jet axis that have not been observed. In reality, the amount of backflowing gas reaching the $z=0$ symmetry plane may be much less because the jets in our simulations are initiated at the origin with full kpc-sized crossections, while in reality young jets are conical with much smaller crossections. We intend to investigate this issue in the future. 
 
     \begin{figure}
\plotone{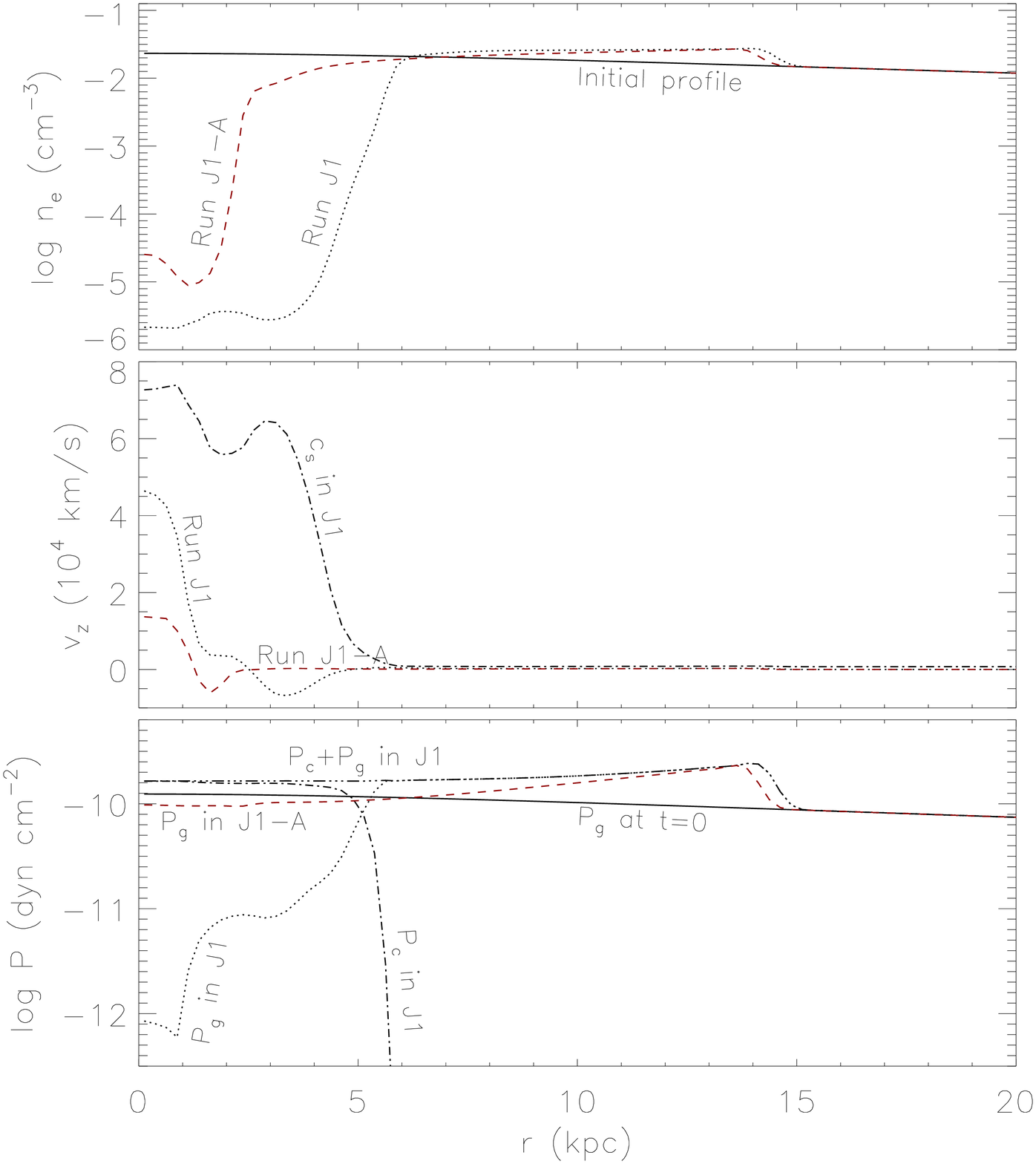}
\caption{Variations of electron number density (top), the $z$-component gas velocity (middle), and pressures (bottom) along the $r$-direction (perpendicular to the jet axis) at $z=10$ kpc for runs J1-A (dashed; red) and J1 (other line types; black) at $t=10$ Myr. The initial gas density and pressure profiles along the $r$-direction at $t=0$ are plotted as solid lines in the top and bottom panels, respectively. The dot-dashed line in the middle panel shows the sound speed in run J1 at at $t=10$ Myr. 
}
 \label{plot4}
 \end{figure} 
 
\begin{figure}
\plotone{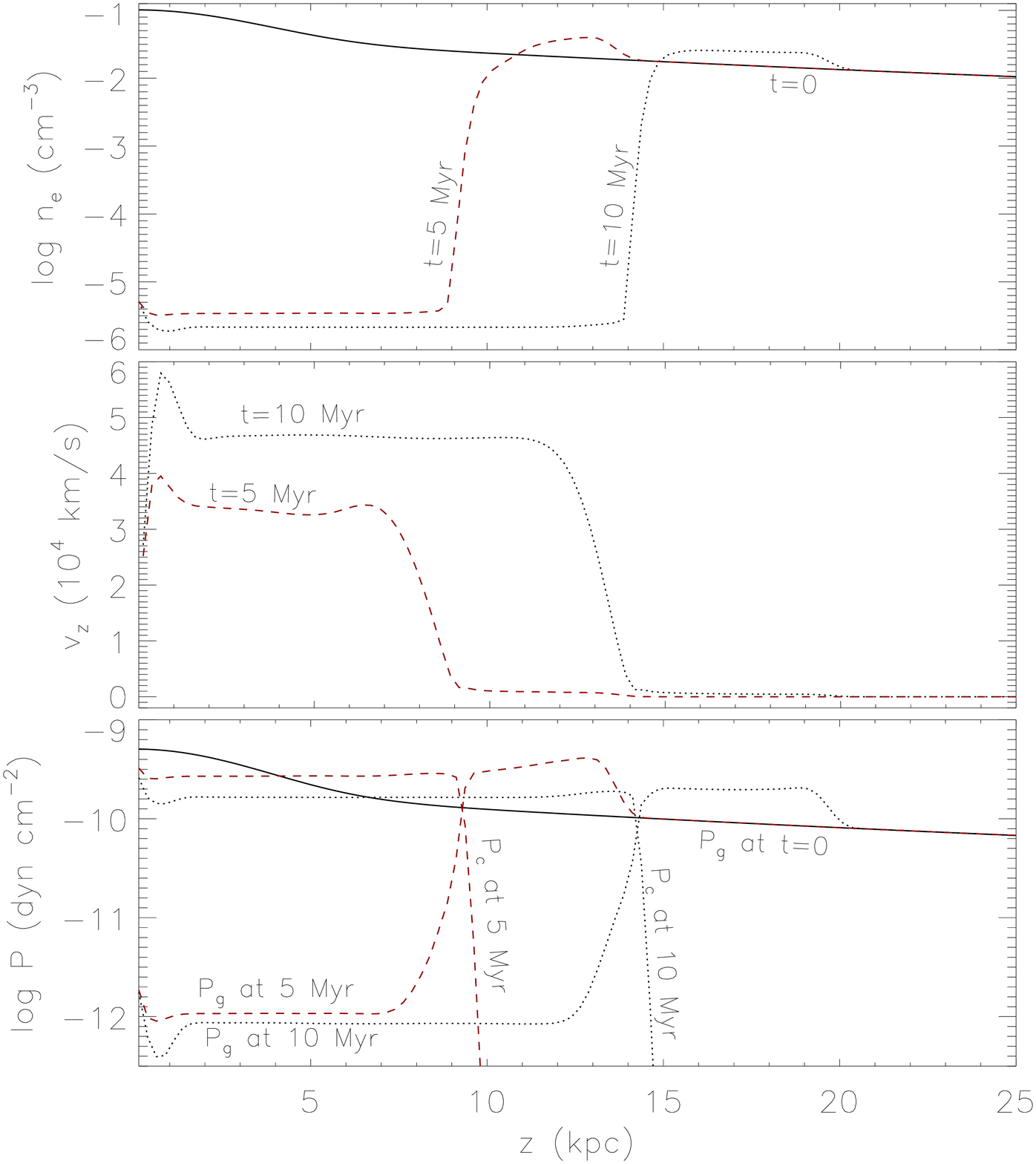}
\caption{Variations of electron number density (top), the $z$-component gas velocity (middle), and pressures (bottom) along the jet axis for run J1 at $t=5$ Myr (dashed) and $t=10$ Myr (dotted). The initial gas density and pressure profiles along the $z$-axis at $t=0$ are plotted as solid lines in the top and bottom panels, respectively. Note that the jet is initialized in ghost zones, which are not plotted.}
 \label{plot5}
 \end{figure} 
 
 \subsection{Why do CR-dominated Jets Work?}
 \label{section:why}

Why are CR-dominated jets successful in creating `fat' X-ray cavities near the cluster center? The CR-dominated jet in run J1 differs from the kinetic-energy-dominated jet in run J0 mainly in two aspects: the dominant CR component and the lower gas density (by two orders of magnitude). To disentangle these two factors, we performed run J1-A, where the initial jet parameters are the same as those in run J0, except that the thermal gas density in the initial jet is much lower ($\eta=10^{-4}$ instead of $\eta=10^{-2}$), as adopted in run J1 (see Table \ref{table1} for the jet parameters and energetics). The bottom panels of Figures \ref{plot1} and \ref{plot2} show the jet evolution in run J1-A, indicating that purely thermal jets with much lower gas densities decelerate much faster. In fact, thermal jets with lower initial gas densities have lower inertia and lower momentum, and are thus easier to decelerate in the ICM. However, unlike the CR-dominated jet in run J1 which creates a fat cavity resembling observed X-ray cavities, the under-pressured J1-A jet creates a strongly elongated cavity. Furthermore, the CR-dominated cavity in run J1 is located at an even smaller cluster-centric distance than that in run J1-A at a same time. Note that the initial jet in runs J1 and J1-A contains the same thermal component, and the difference is that only J1 contains an additional dominant CR component. Figures \ref{plot1} and \ref{plot2} show that the CR-dominated jet in run J1 creates a much wider cavity that experiences a larger ram pressure as it moves away from the cluster center. Consequently, the slower outward velocity of the J1 cavity quickly becomes dominated by buoyancy as observations suggest, while the initial jet momentum may drive the continued outward motion of the narrow cavity in run J1-A for a longer time. Thus, both the very low gas density and the CR pressure help the J1 jet to form a cavity near the cluster center. A high internal pressure from CRs in the J1 jet is required to create a `fat' cavity.

To better understand the formation of fat X-ray cavities by CR-dominated jets, we show the transverse jet structure in Figure \ref{plot4}, which plots transverse variations of electron number density, the $z$-component gas velocity $v_{\rm z}$, and the gas and CR pressures along the $r$-direction (perpendicular to the jet axis) at $z=10$ kpc for runs J1 and J1-A at $t=10$ Myr. Consider first run J1-A (the dashed line in each panel). The transverse density profile clearly shows a bow shock at $r\sim 15$ kpc, which encloses an over-pressured cocoon of shocked cluster gas, a low-density cavity of shocked jet gas, and the jet moving along the $z$ axis. The velocity structure shown in the middle panel of Figure \ref{plot4} indicates that the jet velocity drops quickly away from the jet axis, and even becomes negative at outer regions of the low-density cavity. Low-density cocoon gas with negative $z$-component velocities are backflows from the jet working surface with the ambient ICM. The outer boundary of the low density cavity in J1-A is located at the contact discontinuity between the backflowing cavity material and the shocked cluster gas, across which the gas pressure is continuous, as clearly seen in the bottom panel of Figure  \ref{plot4}. By contrast, for run J1, the low-density cavity is dominated by CR pressure, and the total pressure (including pressures from both the gas and CRs) is also continuous across the outer boundary of the cavity. 

The low-density cavity in jet J1 is much wider than in J1-A and contains thermal gas with much lower densities, as shown in the top panel of Figure \ref{plot4}. Interestingly, the cavity gas density in run J1 ($n_{\rm e}\sim 2$ - $4 \times 10^{-6}$ cm$^{-3}$) is even lower than the density of injected thermal gas at the jet base ($n_{\rm e}\sim  10^{-5}$ cm$^{-3}$ for $\eta=10^{-4}$), which is a signature of the jet expansion due to the initial high-pressure CRs. Due to its extremely low density, the gas within the J1 jet and cavity will be quickly accelerated by any pressure gradient. In particular, the extremely low gas density also ensures a very high sound speed ($c_{\rm s}=\sqrt{(\gamma P_{\rm g}+\gamma_{\rm c} P_{\rm c})/\rho}$), rendering the jet velocity to be subsonic with respect to the jet material and the low-density cavity, as clearly seen in the middle panel. This interesting feature distinguishes very light jets from heavier jets like J0, which is supersonic with respect to both the jet material and its cavity. A direct consequence for ultra-light subsonic jets like J1 is that the total pressure is nearly constant throughout the cavity except near the origin, where the jet is quickly accelerated and expanded (both laterally and radially) due to the pressure gradients produced by the jet injection. This is shown in Figure \ref{plot5}, which plots variations of electron number density, the $z$-component gas velocity, and pressures along the jet axis at $t=5$ Myr (dashed) and $t=10$ Myr (dotted) for run J1. The bottom panels of Figures \ref{plot4} and \ref{plot5} also show that the jet and cavity in run J1 are dominated by the CR pressure.

\begin{figure}
\plotone{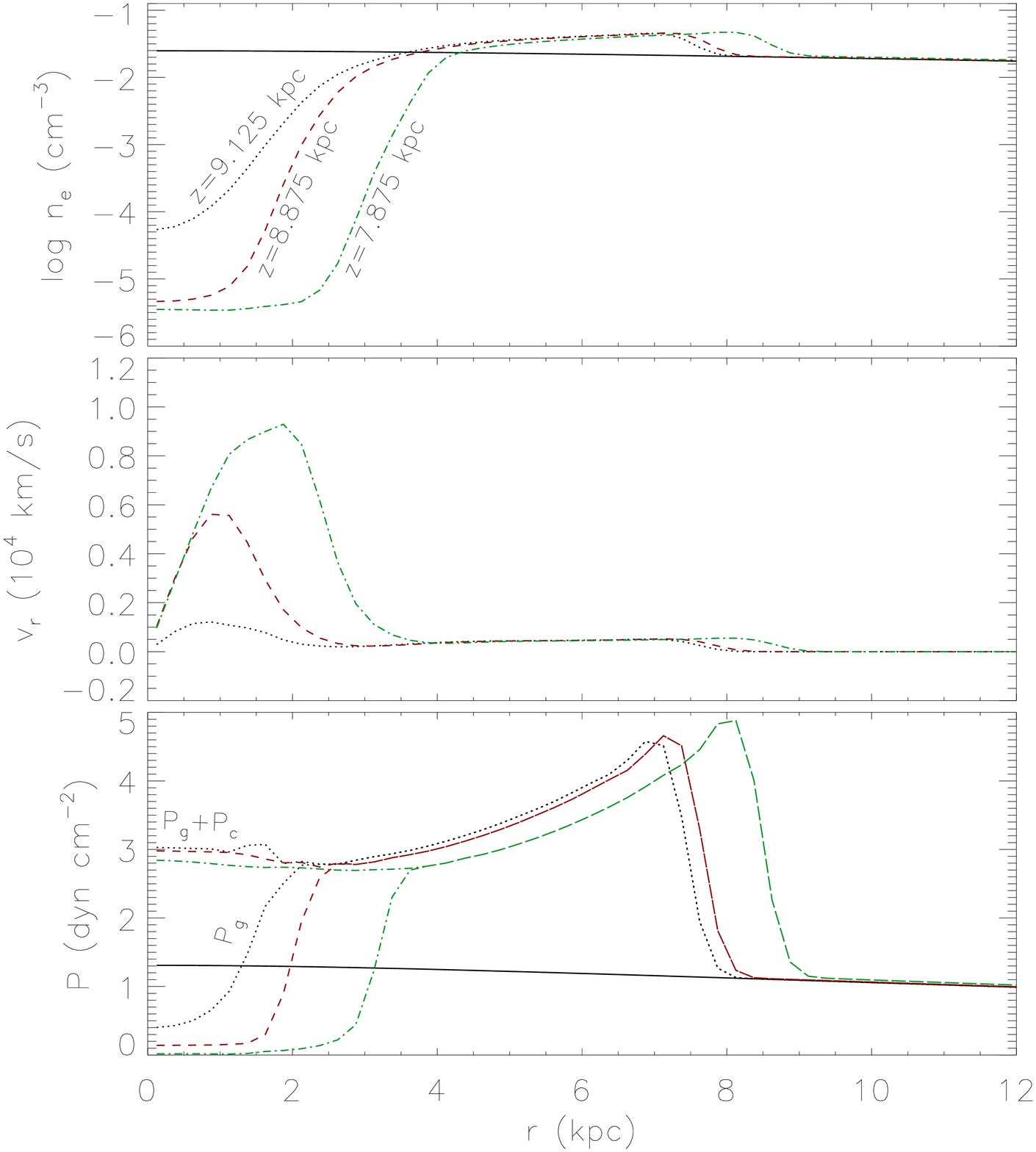}
\caption{Variations of electron number density (top), the $r$-component gas velocity (middle), and pressures ($P_{\rm g}$ and $P_{\rm g}+P_{\rm c}$) along the $r$-direction (perpendicular to the jet axis) for run J1 at $z=9.125$ (dotted lines), $8.875$ (dashed lines), and $7.875$ kpc ( dot-dashed lines) at $t=5$ Myr. The initial gas density and pressure profiles along the $r$-axis at $z=9.125$ kpc are plotted as solid lines in the top and bottom panels, respectively. Near the jet axis, the gas pressure drops significantly in the low density cavity where CR pressure dominates.}
 \label{plot6}
 \end{figure}

The lateral jet expansion, which is essential in forming a fat cavity, also happens in the jet head regions near its working surface with the ICM. Figure \ref{plot6} shows transverse variations of electron number density, the $r$-component gas velocity $v_{\rm r}$, and pressures along the $r$-direction (perpendicular to the jet axis) at $z=9.125$, $8.875$, and $7.875$ kpc at $t=5$ Myr, when the jet at these $z$ values is near the jet head, as clearly seen in the dashed lines in Figure \ref{plot5}. The middle panel of Figure \ref{plot6} shows that the low-density jet and cavity are undergoing significant lateral expansion in these regions, which is driven by the small lateral total pressure gradient in low density gas, as seen in the lower panel. In fact, as the gas and CRs in the jet move quickly down the cluster pressure gradient, the jet becomes slightly over-pressured near its working surface, producing the lateral pressure gradient. The resulting cavity is fatter when the injected jet pressure is larger at the jet base.

For a direct comparison with the kinetic-energy dominated J0 jet, our initial parameters for the CR-dominated J1 jet are chosen so that it has the same energy flux as J0, but mainly in the form of CRs. To this end, we chose the initial CR energy density in the jet to be $e_{\rm jcr}=2.96\times 10^{-9}$ erg cm$^{-3}$, which makes the jet over-pressured at the jet base (the ratio of the jet-to-ICM pressure is about $2$). In principle, the initial jet pressure is determined by physical processes directly related to the central supermassive black hole, and is unlikely to equal to the pressure of central hot ICM. The jet may become over-pressured as it quickly moves down the cluster pressure gradient. Moreover, CRs may be produced within the jet, rendering the jet suddenly over-pressured (e.g., Laing et al. 2006). Run J1 thus represents these cases. 

\begin{figure}
\plotone{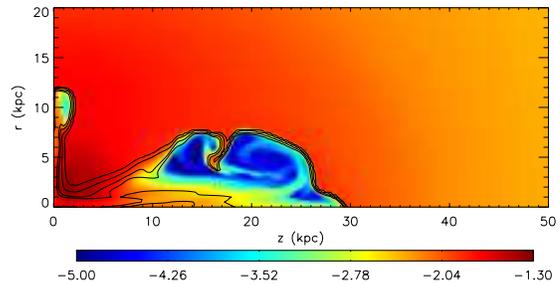}
\caption{Slice of  log $(n_{\rm e}/{\rm cm}^{-3})$ in run J2 at time $t=60$ Myr. Contours show CR energy density $e_{\rm cr}$ in units of $10^{-10}\text{ erg cm}^{-3}$ at four levels $0.01$, $0.1$, $0.5$, $1.0$.}
 \label{plot7}
 \end{figure}

We also consider another run J2 in which the jet is initially in pressure equilibrium with the ambient ICM at its base. The J2 jet differs from J1 only in having a lower CR energy density at the jet base, $e_{\rm jcr}=1.52\times 10^{-9}$ erg cm$^{-3}$. The total jet energy flux in run J2 is thus less than that in run J1 (see Table 1). Figure \ref{plot7} shows a central 2D slice ($50$ $\times$ $20$ kpc) of electron number density in logarithmic scale in run J2 at time $t=60$ Myr. As in run J1, the CR-dominated jet in run J2 produces a fat X-ray cavity near the cluster center. The jet and cavity structure is very similar to that in run J1. In contrast, the jet in run J1-A without CRs produces a strongly radially elongated cavity, suggesting that a minimum amount of CR component is required to produce a fat X-ray cavity near the cluster center.

\begin{figure}
\plotone{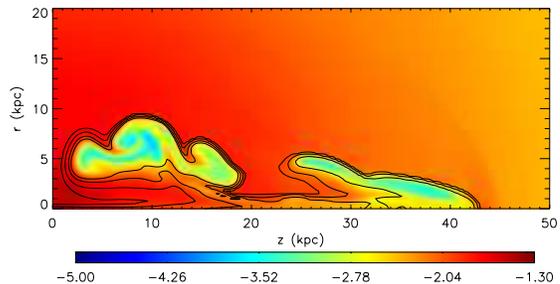}
\caption{Slice of  log $(n_{\rm e}/{\rm cm}^{-3})$ in run J3 at time $t=70$ Myr. Contours show CR energy density $e_{\rm cr}$ in units of $10^{-10}\text{ erg cm}^{-3}$ at four levels $0.01$, $0.1$, $0.5$, $1.0$.}
 \label{plot8}
 \end{figure}

Successful CR-dominated jets discussed above have extremely low initial densities ($\eta=10^{-4}$) and the jet energy flux and pressure are mainly carried by CRs. We have also performed a few simulations of CR-dominated jets with larger gas densities $\eta=10^{-2}$ (i.e., with jet electron number densities $\eta n_0\sim 10^{-3}$ cm$^{-3}$). By increasing the CR pressure within the jet, we find that as expected, the cavity produced rises in the ICM much slower than in run J0, where no CRs are introduced. However, due to the higher jet inertia and momentum, the cavity formed is usually significantly elongated in the jet direction, similar to run J0. This can be seen in Figure \ref{plot8}, which illustrates a central 2D slice ($50$ $\times$ $20$ kpc) of  log $(n_{\rm e}/{\rm cm}^{-3})$ in a typical run J3 at time $t=70$ Myr (see Table 1 for simulation parameters in this run). Although the backflowed gas forms a fat cavity that rises slowly in the ICM (mainly due to buoyancy), a large fraction of the jet material, in particular the jet head, still moves quite fast, driven by the large initial jet momentum. The cavity formed is significantly radially elongated, strikingly different from the cavity produced in run J1 (Figures \ref{plot1} and \ref{plot2}). Thus our preliminary calculations suggest that observed X-ray cavities are likely to have been produced by CR-dominated light jets with $\eta \ll 0.01$. Jet simulations involve a large parameter space, and we leave a more thorough parameter study to future work, which we anticipate will also be possible to explain observed X-ray cavities and radio bubbles spanning a large range of size, cluster-centric distance, and age \citep{mcnamara07}. 

The main dynamical effect of CRs is their contribution to the ICM pressure while introducing negligible inertia. The same effect may also be achieved with a purely thermal jet by proportionally increasing the gas temperature while extremely reducing the gas density. However, to reach a jet-to-ICM density ratio of $\eta=10^{-4}$, the jet temperature should be at least $\sim 10^4$ keV (i.e. $T \gtrsim 10^{11}$ K), and gas with such high temperatures is already mildly relativistic [electron gamma-factor $\gtrsim 30(k_{\rm B}T/10^{4}$ keV$)$]. Radio synchrotron emission from FR I jets and some X-ray cavities indicates a significant population of CR electrons with higher energies (gamma-factor of $\sim 10^{3}-10^{5}$). In our model, the dynamics of CRs is described by $e_{\rm c}$, the integrated energy density over the CR energy distribution. Our model is applicable to relativistic CRs with any energy spectrum or any content (i.e., any combination of CR electrons and protons). A proper consideration of these details is necessary to calculate the radio emission from CRs and this is an important problem for the future.

 \section{Conclusion and Implication}
\label{section:conclusion}

AGN feedback has been frequently invoked to explain the cut-off of the galaxy luminosity function at the bright end and the suppression of cooling flows in giant elliptical galaxies, galaxy groups, and clusters. Many X-ray deficient cavities detected by high-resolution X-ray observations provide strong evidence for AGN feedback in galaxy clusters. Most cavities are roughly spherical or slightly elongated in the tangential direction, and those detected are commonly located near cluster centers (offset by a few to tens kpc), having spatial sizes of a few to tens kpc. However, in numerical simulations it has been difficult to reproduce such fat cavities. Hot purely thermal jets frequently adopted in previous simulations usually penetrate deeply and rapidly through the ICM, forming radially-elongated cavities at much larger radii.

In this paper, for the first time, we present numerical simulations of CR-dominated AGN jets and follow their dynamical evolution as they form cavities in the ICM. Using two-dimensional axisymmetric calculations, we show that CR-dominated jets naturally create fat cavities near cluster centers, resembling observed cavities very well. 
The key feature of successful CR-dominated jets in our model is that they are also extremely light -- the jet-to-ICM density ratio is $10^{-4}$ in our main run J1. The low inertia and momentum in these jets ensure that they are quickly decelerated in the ICM. The energy fluxes of these jets are dominated by CR energy, and the CR pressure drives the lateral jet expansion, which assists the jet deceleration and is essential in forming fat lobes. The resulting lobes are mainly formed by jet backflows, and supported by the CR pressure, which also displaces ambient thermal gas, naturally creating X-ray deficient cavities filled with synchrotron emitting CR electrons. The initial low jet momentum is quickly transferred to the ambient post-shock cluster gas, and the X-ray cavities then rise buoyantly in the ICM.

Our scenario in which X-ray cavities evolve from CR-dominated AGN jets is supported by observations of CR-filled X-ray cavities in elliptical galaxies, galaxy groups and clusters which are often associated with synchrotron-emitting radio jets and spatially coincident with radio bubbles \citep{mcnamara07}. High-resolution multi-frequency radio observations of FR I jets, which are probably responsible for forming X-ray cavities, indicate that these jets decelerate rapidly and produce strong synchrotron emission (flaring) at typical distances of a few kpc from central nuclei (e.g., \citealt{laing06}). These observations thus also point toward a significant CR electron population in FR I jets on kpc scales. 

Our calculations of cavity formation imply that, to create observed fat X-ray cavities near cluster centers, AGN (FR I) jets must be (1) very light, containing a very small thermal component, and (2) dominated by CR pressure. We thus derive new constraints on the composition of FR I jets solely from their ability to create fat X-ray cavities (radio bubbles) as observed in elliptical galaxies, groups and clusters of galaxies. The thermodynamical impact of CR-dominated jets on the ICM, e.g., their efficiency in heating cooling flows and spreading metals, may be different from that of purely thermal jets, and this will be investigated in future studies.    

\acknowledgements
We thank the anonymous referee for a positive and detailed report, which was very informative and helpful. Studies of CR-dominated AGN jets and the evolution of hot cluster gas at UC Santa Cruz are supported by NSF and NASA grants for which we are very grateful.


\label{lastpage}

\end{document}